\documentclass[conference,compsoc]{IEEEtran}
\ifCLASSOPTIONcompsoc
  \usepackage[nocompress]{cite}
\else
  \usepackage{cite}
\fi
\ifCLASSINFOpdf
\else
\fi

\usepackage{graphicx}
\usepackage{makecell}
\usepackage{amsmath}
\usepackage{xcolor}
\usepackage{booktabs}
\usepackage{multirow}
\usepackage{multicol, blindtext}
\usepackage{booktabs}
\usepackage{hyperref}

\begin{document}
%
\title{Baseline Pruning-Based Approach \\ to Trojan Detection in Neural Networks}

\author{\IEEEauthorblockN{Peter Bajcsy\IEEEauthorrefmark{1} and Michael Majurski }
\IEEEauthorblockN{
Information Technology Laboratory \\
National Institute of Standards and Technology \\
100 Bureau Drive. Gaithersburg, MD 20899 \\
\IEEEauthorrefmark{1}Email: peter.bajcsy@nist.gov}
}


%


\maketitle

\begin{abstract}
This paper addresses the problem of detecting trojans in neural networks (NNs) by analyzing systematically pruned NN models.
Our pruning-based approach consists of three main steps. First, detect any deviations from the reference look-up tables of model file sizes and model graphs. 
Next, measure the accuracy of a set of systematically pruned NN models following multiple pruning schemas. Finally, classify a NN model as clean or poisoned by applying a mapping between accuracy measurements and NN model labels. This work outlines a theoretical and experimental framework for finding the optimal mapping over a large search space of pruning parameters. Based on our experiments using Round 1 and Round 2 TrojAI Challenge datasets, the approach achieves average classification accuracy of $69.73\:\%$ 
and $82.41\:\%$ respectively with an average processing time of less than $60\:s$ per model. 
For both datasets random guessing would produce $50\:\%$ classification accuracy.
Reference model graphs and source code are available from GitHub.
\end{abstract}


%
\IEEEpeerreviewmaketitle

\section{Introduction}

\begin{figure}
\includegraphics[scale=0.5]{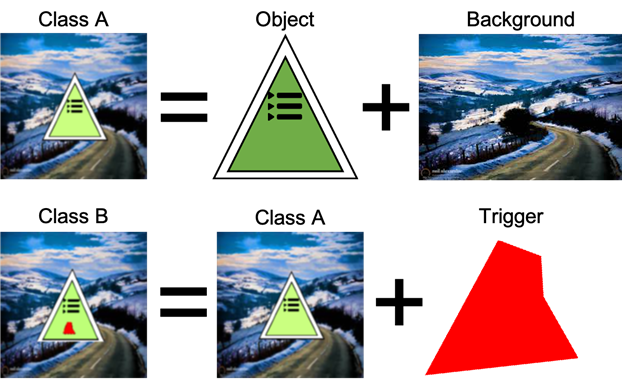}
  \centering
  \caption{Illustration of injecting a polygon trojan (trigger) to a traffic sign region that causes the shift in classification from class A to class B.}
  \label{fig:01}
\end{figure}

This work addresses classifying neural network (NN) models into two classes: (1) models trained without trojans (clean) and (2) models trained with trojans (poisoned).
Trojans in NNs are defined as triggers inserted into the inputs that cause misclassification into a class (or classes) unintended by the design of the model \cite{Bajcsy2020}. 
For example,  trojans can be polygons inserted as innocuous objects (triggers) into traffic sign images (foreground) to change the classification result as shown in Figure \ref{fig:01}. 
Such triggers have been used to generate the datasets for multiple rounds of the Intelligence Advanced Research Projects Agency (IARPA) challenge \cite{IARPA2020}. 

The overarching motivation for designing trojan detection algorithms is the defense against a variety of adversarial attacks during NN training. NN training might be outsourced to a third party with unknown malicious intent. It might also leverage NN models pre-trained by an unknown untrusted third party. In many life-critical applications, such as self-driving cars or medical diagnoses, deployment of NN models depends on establishing trust in model performance. To build that trust, trojan detection algorithms must operate on a variety of model architectures, with limited prior knowledge 
about the model, and for a wide range of trojan types.
Our work is motivated by the need to establish a baseline approach for new and innovative algorithms tested on the IARPA challenge datasets.
In addition, we are motivated to lower the trojan detection computational requirements to the level of a simple phone app.

\begin{figure}
\includegraphics[
  width=8cm,
  keepaspectratio,
]{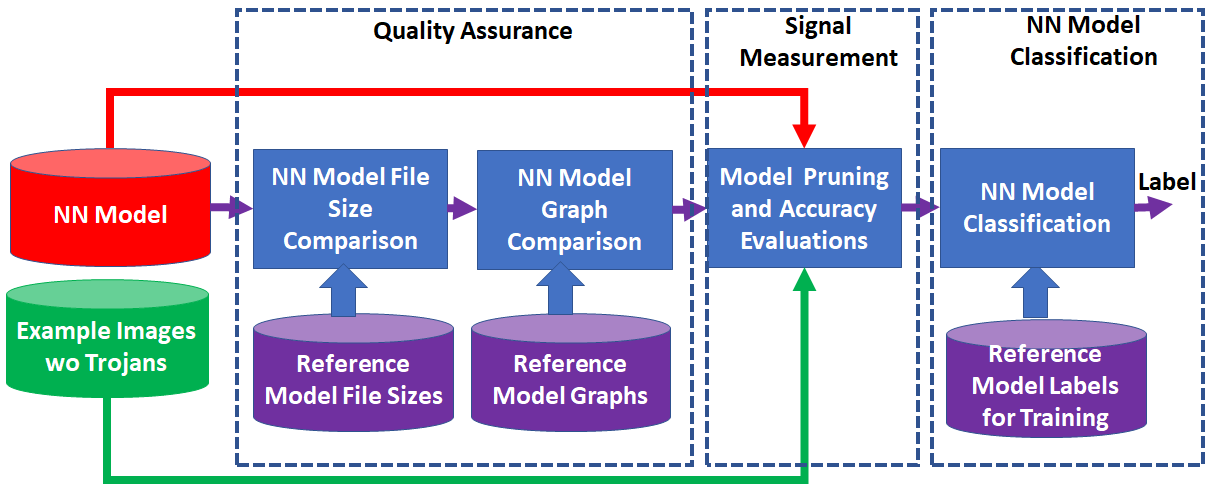}
  \centering
  \caption{Overview of NN model classification workflow}
  \label{fig:02}
\end{figure}

The goal of this work is to design a baseline approach for detecting (a) possible tampering with the reference model architecture (changing task-specific NN architecture called a reference model)
and (b) the presence of trojans in a spectrum of architectures. 
Our approach is illustrated in Figure \ref{fig:02}.
The ``quality assurance'' computations in Figure \ref{fig:02} are based on our prior knowledge about model files and architecture graphs in order to detect deviations from reference models. 
The ``signal measurement'' computations in Figure \ref{fig:02} focus on measuring accuracies of systematically pruned models. Finally, the ``NN model classification'' computations derive and apply a mapping between accuracies of pruned models and labels denoting the presence of an embedded trojan. The main challenges lie in estimating the optimal mapping, in collecting signal measurements within a time limit, and in making the mapping robust to many architectures and to complex trojan characteristics.  

Our contributions lie in the design of a baseline trojan detection approach that
\begin{itemize} 
\item leverages well-established filter pruning approaches and their existing implementations (provides a baseline), 
\item evaluates multiple pruning, ranking, and sampling methods into model pruning (includes optimization),
\item collects model accuracy measurements over a wide spectrum of architectures and with varying number of input images (delivers robustness), and
\item includes classification accuracy and execution speed tradeoffs into the trojan detection design (measures scalability). 
\end{itemize}

\section{Related Work}

The design of trojan detection algorithms is a relatively new area of research. According to the statistics derived from the publications listed at \cite{relevantPapers2020} in 2020, two related publications appeared in Arxiv before 2017, eight in 2017, 15 in 2018, 31 in 2019, and 57 in 2020. The research interest increased as IARPA and Defense Advanced Research Projects Agency (DARPA) announced the TrojAI~\cite{IARPA2020} and 
Guaranteeing AI Robustness Against Deception (GARD)~\cite{Siegelmann2019} programs in 2019.  With more research efforts invested into designs of trojan detectors \cite{Xu2019, Jha2019, Erichson2020}, there is a need to establish a baseline method that is simple, but generally applicable, and provides results that are better than a chance~\cite{baseline2020}. 

Model pruning approaches have been very popular in the past NN research
\cite{Blalock2020,BabakHassibi1992,Han2015,Hu2016,Li2017,Liu2018a}. The past model pruning objectives were focused on reducing file size and computations needed for storing, training and inferencing models 
\cite{Han2015,anwar2015structured,Hu2016,See2016,Li2017,mariet2017diversity,Molchanov2017,Ye2018}.
The plethora of pruning approaches was documented in a survey of 81 papers that led the authors in \cite{Blalock2020} to design a framework for evaluations of pruning methods. Such a wide use of the model pruning approach motivated us to leverage this approach for a design of the baseline trojan detector. In addition, model capacity, efficiency, and model pruning were mentioned as factors and a possible solution that can increase robustness and resiliency to attacks \cite{Liu2018,Liu2018a}.

Our survey of available GitHub pruning-based solutions~\cite{pruninglink2020} highlighted the existing challenges in terms of the limited number of supported model architectures, long execution times, and dependencies on outdated libraries. For example, the  GitHub implementation from \cite{Molchanov2017} is applicable to VGG16 architectures and has been adapted to ConvNet, AlexNet , ResNet18, InceptionV3, and ResNet50 in limited settings \cite{pruninglink2020}. There is no pruning implementation that would work with the 22 model architectures presented in the TrojAI challenge. Thus, our work could only partially leverage the GitHub implementation linked from \cite{Li2017}.
 
\section{Methods}
\label{sec:methods}

\underline{Classification Problem:}
The problem can be formulated as follows: 
\\
Classify a set of NN models $M$ as clean or poisoned such that the classification is robust to 
\begin{itemize} 
\item architecture, 
\item the number of provided sample images without trojans, and
\item trojan type; 
\end{itemize} 
while execution time is limited on a variety of computational hardware. 
\\
Formally, given the following inputs:
\begin{itemize} 
\item a set of clean images $D_i$ that represent samples from each predicted class $C_l \in C$; 
\item 
a model $M_i \in M$ of an architecture $G_n \in G$ that predicts $|C|$ classes
\item a corresponding label for each model $M_i$: 
	 \begin{itemize} 
	 \item $L_i= 0 \rightarrow$ clean or Trained without Trojan,
	 \item $L_i=1 \rightarrow$ poisoned or Trained with Trojan,
	\end{itemize}  
\end{itemize} 
Classify the model $M_i$ as either clean or poisoned
while minimizing the trojan detection error within an allocated execution time $T_i \leq T_{max}$ on a variety of computational platforms.  Note: $|C|$ refers to the number of classes (cardinality of the set of labels $C$).

\underline{Pruning-based Approach:}
To solve the classification problem, we introduced quality assurance (QA) based classification criteria and designed a supervised pruning-based classifier. The QA-based classification assumes reference measurements about file size and model graphs are known. The pruning-based classifier assumes that a trojan is encoded in convolutional filters. Thus, one can discriminate NN models into clean and poisoned categories by systematic pruning of convolutional filters across all layers, measuring accuracies of pruned models $\Vec{A_i}$, and estimating some function $f(\Vec{A_i}) \rightarrow L_i$. 

The pruning-based approach is characterized by the search for an optimal mapping function $f(\Vec{A_i})$ and optimal parameters $\theta_{opt}{(G_n, D)}$ used for computing the vector of pruned model accuracies 
$\Vec{A_i}$. The set of optimal parameters $\theta_{opt}{(G_n, D)}$ is specific to each NN architecture $G_n$ and depends on available clean images $D_i \in D$. The optimization task can be formally defined as follows:
\\
Given the pruning-based approach and the following inputs:
\begin{itemize} 
\item a NN model $M_i{(G_n, C)} \in M$, 
\item a set of clean images $D_i{(C)}$, and 
\item a clean or poisoned label for each model $L_i$; 
\end{itemize} 
Find an optimal configuration of parameters $\theta_{opt}{(G_n, D)}$ for each model architecture $G_n \in G$ that minimizes the NN classification error $\mathcal{L}_{i}^{error}$ subject to allocated execution time $\mathcal{L}_{i}^{exec}$ per NN model as shown in Equation \ref{eq:01}.
\begin{equation} \label{eq:01}
\begin{split}
\min_{\theta(G_n,D)}{\sum_{i=1}^{|M(G_n)|}
 \frac{1}{|M(G_n)|}*\mathcal{L}_{i}^{error}(\theta(G_n,D))
 }\\
\textrm{subject to} \quad
\mathcal{L}_{i}^{exec}(\theta(G_n,D))  \leq 1
\end{split}
\end{equation}
\noindent  where 
$|M(G_n)|$ is the number of NN models of the $G_n$ architecture type and  $\theta(G_n,D)$ is a set of algorithmic configurations evaluated for each NN architecture type $G_n$. 
The term for classification error $\mathcal{L}_{i}^{error}$ is defined as $\mathcal{L}_{i}^{error} = 1.0 - \mathcal{L}_{i}^{AC}$ defined in Equation \ref{eq:02} or as a cross entropy (CE) loss $\mathcal{L}_{i}^{CE}$ according to Equation \ref{eq:02b} (see also \cite{trojai-data2020}).
In these equations, $\Vec{A_i}=\Vec{A_i}{(M_i, \theta(G_n,D))}$ is a vector of accuracy measurements over pruned models, $f(\Vec{A_i})$ is the probability of predicting a poisoned model, $\lfloor \; \rceil$ denotes rounding to the nearest integer, and $\big[ \; \big]$ is the Iverson bracket.
The term for execution time $\mathcal{L}_{i}^{exec}$ is defined as a percentage of maximum allocated execution time $T_{max}$ according to  Equation \ref{eq:03}. 
\begin{equation} \label{eq:02}
\mathcal{L}_{i}^{AC} = 
\Big[ L_i = \lfloor f(\Vec{A_i}) \rceil  \Big] 
\end{equation} 

\begin{equation} \label{eq:02b}
\mathcal{L}_{i}^{CE}=
- (L_i*ln( f(\Vec{A_i}) ) + (1-L_i)*ln(1-f(\Vec{A_i}) )
\end{equation} 

\begin{equation} \label{eq:03}
\mathcal{L}_{i}^{exec}=
\frac{T_i}{T_{max}} 
\end{equation}

\underline{Pruning configurations:}
The space of pruning configurations can be characterized by six parameters: $\theta(G_n, D)=\{PM, SM, RM, p, |S|, |D| \}$.
Pruning methods $PM$ consist of \{\texttt{Remove}, \texttt{Reset}, \texttt{Trim}\}, 
sampling methods $SM$ can be \{\texttt{Random}, \texttt{Uniform}, \texttt{Targeted}\}, 
ranking methods $RM$ include \{$l_1$, $l_2$, $l_{\infty}$, \texttt{stdev} (standard deviation)\}, and
a sampling probability  $p$ can be in general any real value $p \in (0,1) \in \mathcal{R}$ per NN layer. The number of evaluated pruned models per configuration $|S| \in \mathcal{Z}^{>0}$ and
the number of used evaluation images $|D| \in \mathcal{Z}^{>0}$ can be any integer value smaller than the number of all available clean images in $D_i$. Note that we excluded the pruned module type as a pruning parameter as we focus on trojan feature formation in convolutional layers represented by Conv2D and BatchNorm modules. We also excluded the run-time parameters (software and hardware) as they could be optimized based on the requirements for $T_{i}$.

\begin{figure}
\includegraphics[
  width=8cm,
  keepaspectratio,
]{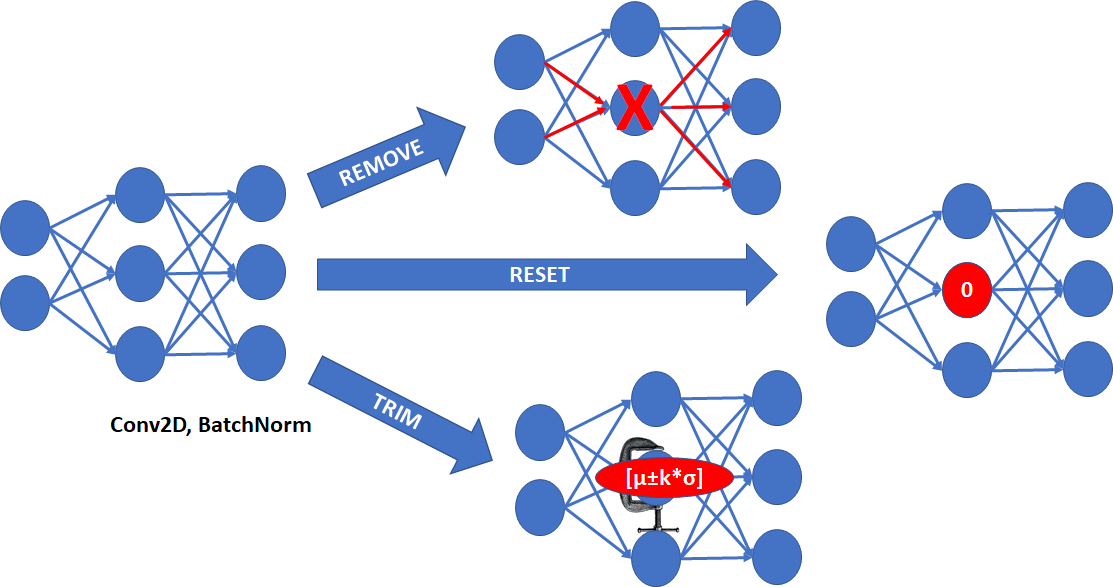}
  \centering
  \caption{Differences between \texttt{Remove} (top), \texttt{Reset} (middle), and \texttt{Trim} (bottom) pruning methods. }
  \label{fig:04}
\end{figure}

\begin{figure}
\includegraphics[
  width=8cm,
  keepaspectratio,
]{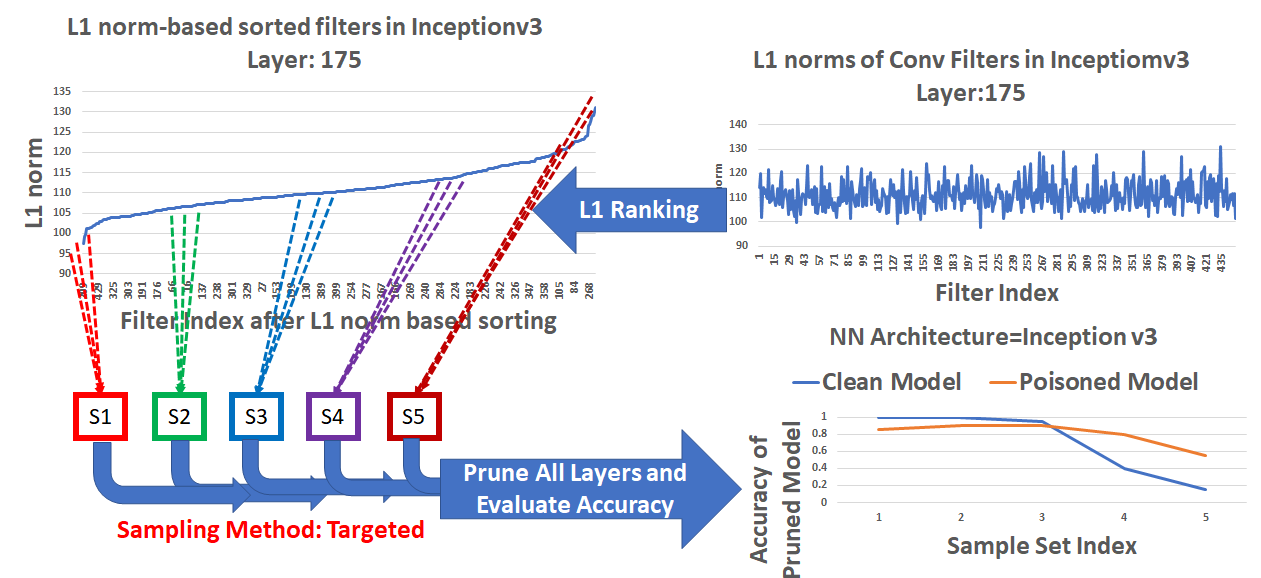}
  \centering
  \caption{Illustration of targeted sampling method and $l_1$ ranking method. }
  \label{fig:05}
\end{figure}

The differences between pruning methods in $PM$ are illustrated in Figure \ref{fig:04}. The \texttt{Remove} method completely removes the convolutional filter and re-connects inputs and outputs. The \texttt{Reset} method sets all filter coefficients to zero and the \texttt{Trim} method clamps the coefficients to the mean $\pm k*stdev$, where the mean and stdev are computed from the convolutional filter coefficients, and $k \in (0,1]$. The sampling methods differ in choosing the set of filters for pruning. Figure \ref{fig:05} shows \texttt{Targeted} sampling method applied after $l_1$ norm was used to rank all filters in one layer (Inception v3 architecture, Layer 175). In Figure \ref{fig:05}, $l_1$ norm is applied to all convolutional filters (top right) and the filters are sorted accordingly (top left). \texttt{Targeted} sampling method selects $|S|=5$ sample sets of filters that are pruned (bottom left). For each of the $|S|=5$ pruned models, the model accuracy is evaluated using $|D|=10$ clean example images. 
Figure \ref{fig:05} (bottom right) shows an example of accuracies measured over $S1, S2, S3, S4$ and $S5$ pruned models for clean and poisoned models.  While \texttt{Targeted} sampling selects contiguous filters from a sorted list, the \texttt{Uniform} sampling method chooses uniformly distributed filters after ranking them. The \texttt{Random} sampling method selects filters randomly and the sampling is repeated $|S|$ times. 

\underline{Reduction of Search Space:}
The challenge of the pruning-based approach applied to convolutional filters on NN models lies in the cost of searching the space of all possible pruned models and pruning methods per architecture $G_n$. 
Theoretically, the number of possible pruned models is $\prod_{j=1}^{|L(G_n)|}(2^{|F_j|} - 1)$ for a NN architecture $G_n$ that consists of $|L|$
convolutional layers with a varying number of convolutional filters $|F_j|$ within each layer. 
We assume that 
\emph{the significance of a convolutional filter to class predictions is related to the norm of the filter coefficients} \cite{Hu2016, Li2017, Ye2018}. 
 Thus, the number of pruned models can be reduced to $\prod_{j=1}^{|L(G_n)|}|F_j|$ by ranking filters; therefore ranking methods are included in the pruning configurations. Unfortunately, there is no theory nor guidelines about how to rank NN convolutional layers based on their influence on the output \cite{Ye2018}. 
 Thus, in order to reduce the search space, we assumed that all layers are equally significant to class predictions and applied the same sampling probability $p$ of removed filters to all layers. 

\begin{figure}
\includegraphics[
  width=8cm,
  keepaspectratio,
]{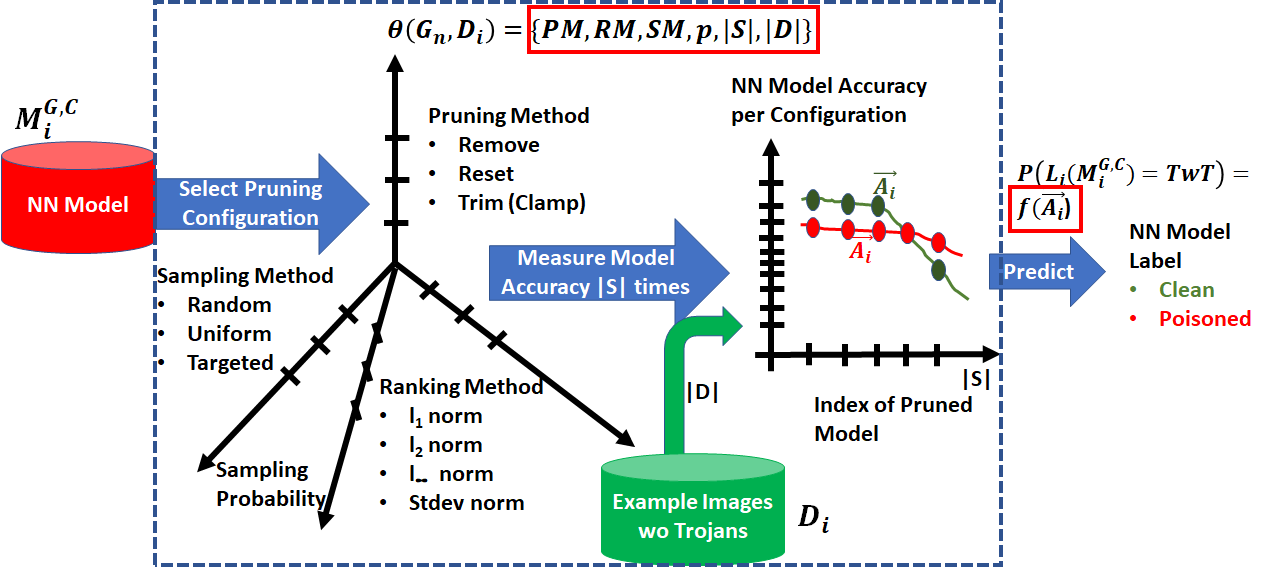}
  \centering
  \caption{Pruning configurations generating measurements.}
  \label{fig:03}
\end{figure}

The challenge of optimizing the pruning-approach parameters l
ies in the additional cost of evaluating all possible parameter configurations. Theoretically, the space of parameters is infinite as it consists of all pruning configuration parameters $\theta_n$ per architecture and all models for the functional mapping $f$. The pruning configuration space is illustrated in Figure \ref{fig:03} with six parameters $\theta(G_n, D)=\{PM, SM, RM, p, |S|, |D| \}$ and one unknown classifier function $f(\Vec{A_i})$.
To reduce the search space, we first restricted the function $f(\Vec{A_i}{(M_i, \theta_n)})$ to a multiple linear regression. The mathematical expression is shown in Equation \ref{eq:04}. The coefficients are derived by using the pairs of accuracy vectors $\Vec{A_i}$ and labels $L_i$. 
\begin{equation} \label{eq:04}
f(\Vec{A_i}{(M_i, \theta_n)}) = 
b_0+\sum_{k=1}^{|S|} b_k * A_{i,k}{(M_i, \theta_n)}
\end{equation}
\noindent  where $|S|$ is the size of vector $\Vec{A_i}$ and $b_k$ are the coefficients derived from a linear regression fit.

Next, we decomposed the configuration parameters $\theta_n$ into two disjoint subsets $\theta_{n}^{error}=\{PM, SM, RM, p\}$ and $\theta_{n}^{exec}=\{|S|, |D|\}$. The split of the parameters is based on our observations that the number of pruned models $|S|$ and the number of images to evaluate each pruned model with $|D|$ are the key contributors to increasing classification time per model. This parameter decomposition allows us to lower the search cost by first optimizing the four parameters in $\theta_{n}^{error}$ for fixed low values in $\theta_{n}^{exec}$, and then by completing the optimization of the two parameters in $\theta_{n}^{exec}$ with fixed optimal values in $\theta_{n}^{error}$.
 
Finally, we reduce the search space by introducing relationships between $p \in (0,1) \in \mathcal{R}$ and 
$|S| \in \mathcal{Z}^{>0}$ parameters under two assumptions: (1) At least one convolutional filter per layer must be removed in each pruned model and therefore the layer with the smallest number of filters defines the sampling probability as $p = 1/\min_{j \in [1, |L|]}{|F_j|}$. (2) Each filter must be removed at least once in the set of $|S|$ pruned models and therefore $p = k/|S|$, where $k$ is a multiplier defining how many times the same filter could be removed in a set of $|S|$ pruned models yielding  $\Vec{A_i}$.

\section{Experimental Results}

The quality assurance and measurements were implemented in Python using the PyTorch and sklearn libraries. The code, installation instructions, and the reference model architecture graphs are available from GitHub~\cite{Bajcsy2021}.
Next, we summarize the input datasets, quality control and performance results.

\subsection{Input Datasets}
TrojAI challenge datasets are described at \cite{trojai-data2020}. Given the notation in Section \ref{sec:methods}, the Round 1 dataset can be characterized by the number of models $|M|=1000$, the number of architectures $|G|=3$ with $G=$\{ResNet50, InceptionV3, DenseNet121\}, the number of predicted classes $|C|=5$, the number of clean images per class $|D|=100$, and $50:50$ split of labels $L_i$ between clean and poisoned.

Similarly, the Round 2 dataset is described by the number of models $|M|=1104$, the number of architectures $|G|=22$, the number of predicted classes randomly varying $|C|=10 \pm5$ or $|C|=20\pm5$, the number of clean images per class randomly varying $|D|=|C|*10$ or $|D|=|C|*20$, and $50:50$ split between clean and poisoned labels $L_i$. The datasets are summarized in Table~\ref{tab:01}.

\begin{table}[]
\caption{Summary of Input Datasets}
\label{tab:01}
\centering
\begin{tabular}{|c|c|c|c|c|}
\hline
Inputs &  $|M|$ & $|G|$ & $|C|$  & $|D|$  \\  \hline
 Round 1 &  1000 &  3 & 5  &  $|C|*100$  \\ \hline
 Round 2 &  1104 &  22 & $10 \pm5$ or $20 \pm5$  & $|C|*10$ or $|C|*20$  \\ \hline
\end{tabular}
\end{table}

\subsection{Quality Assurance}
The input datasets were processed to compute the average and standard deviation of model file size per architecture. The variations in model file sizes are due to the use of PyTorch (version 3.4 and up) and its dependency on the Python Pickle library \cite{pickle2020} 
for data format optimizations and for saving models as serialized objects to disk. 
As a sanity check, by analyzing the model file sizes and model clean/poisoned labels, we confirmed that model file sizes and their variations do not predict clean or poisoned labels.

For trojan detection, we extracted diagrams of abstract model graphs from the Round 1 and Round 2 datasets using the Graphviz library \cite{graphvis2020}. 
The reference graphs can be found in the reference\_data folder of the GitHub repository~\cite{Bajcsy2021} and are used for detecting graph deviations. 
  
\subsection{Performance Results}

All performance benchmarks were collected on a desktop running Ubuntu 18.04, with 8 CPU cores (Intel(R) Xeon(R) Silver 4114 CPU @ $2.20\;\textmd{GHz}$), and $192\;\textmd{GB}$ RAM. The implementation only utilizes CPU resources.

The evaluations over the set of parameters $\theta_n$ were skewed towards $SM=\texttt{Targeted}$ and $RM=l_1$. 
The \texttt{Targeted} sampling method is expected to outperform \texttt{Uniform} and \texttt{Random} sampling methods by design because we anticipate two distinct trends of accuracy values in the vectors $\Vec{A_i}$ for clean and poisoned models (see Figure \ref{fig:05}, bottom right). As the evaluation metric, we used classification accuracy and average cross entropy loss over all models in each dataset. 

\underline{Round 1 dataset:}
We evaluated 31 pruning configurations for 286 DenseNet121 models, 395 ResNet50 models, and 319 InceptionV3 models in 254 h of compute time. The 31 evaluations per NN model have the following distribution of configuration parameters:

\begin{equation}
\small
\begin{aligned}
PM &= \{\texttt{Remove} (14x), \texttt{Trim} (12x), \texttt{Reset} (5x)\}\\ 
SM &= \{\texttt{Targeted} (27x), \texttt{Uniform} (2x), \texttt{Random} (2x)\}\\ 
RM &= \{l_1 (19x), l_{\infty} (1x), \texttt{stdev} (11x)\}\\ 
p &\in [0.075, 0.9] \\
|S| &= \{5 (20x), 10 (4x), 15 (7x)\}\\
|D| &= \{10 (27x), 20 (1x), 30 (1x), 40 (1x), 100 (1x)\}\\
\end{aligned}
\end{equation}

\begin{figure}
\includegraphics[
  width=8cm,
  keepaspectratio,
]{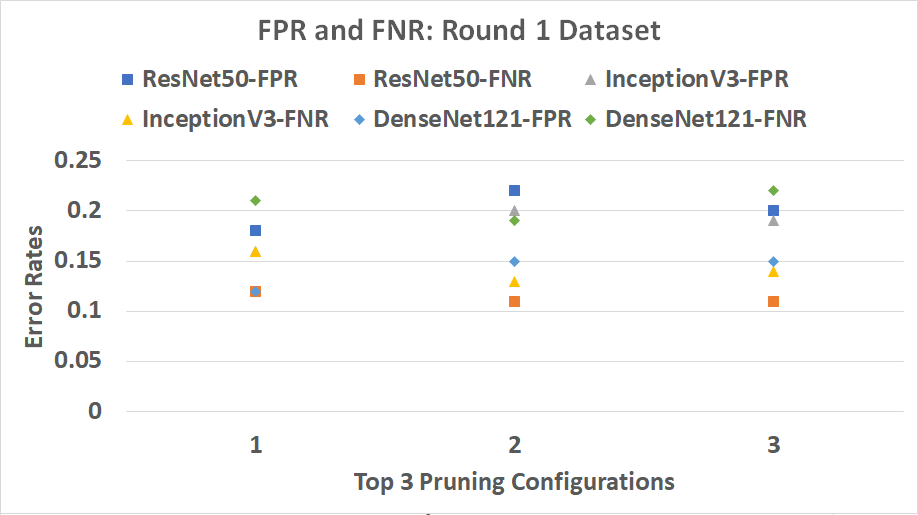}
  \centering
  \caption{False positive (FP) and false negative (FN) error rates from the top three parameter configurations for Round 1 dataset sorted by cross entropy loss. }
  \label{fig:07}
\end{figure}

The sampling probability $p$ was selected based on the assumptions about pruning filters and explored for a wide range of values for the \texttt{Trim} pruning method. The evaluations are staged first for the values of $|S|=5$, $|D|=10$, and then for other values.
We concluded that the smallest classification errors were for ResNet50: $27.85\:\%$, for InceptionV3: $30.09\:\%$, and for DenseNet121: $32.87\:\%$ (average of the three is $30.27\:\%$). When sorted by average cross entropy loss, the smallest values were for ResNet50: $0.5169$, for InceptionV3: $0.5969$, and for DenseNet121: $0.6251$ (average of the three is $0.5796$), where the value of $0.6931$ corresponds to random guessing.
Figure \ref{fig:07} presents the distribution of false positive and false negative error rates in the top three configurations sorted by average CE loss. For these top results, the parameter distribution is skewed towards $PM=\texttt{Remove}$ (6x), $SM=\texttt{Targeted}$ (7x), $RM=l_1$ (9x), and $p=0.02$ (3x), $|S|=15$ (6x), and $|D|=10$ (9x).

\begin{figure}
\includegraphics[
  width=8cm,
  keepaspectratio,
]{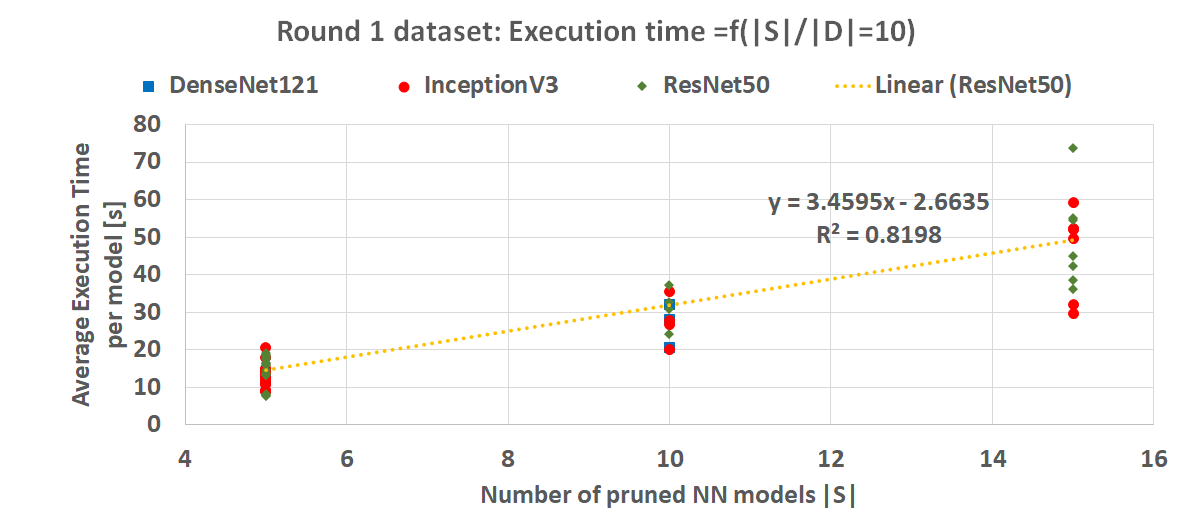}
\includegraphics[
  width=8cm,
  keepaspectratio,
]{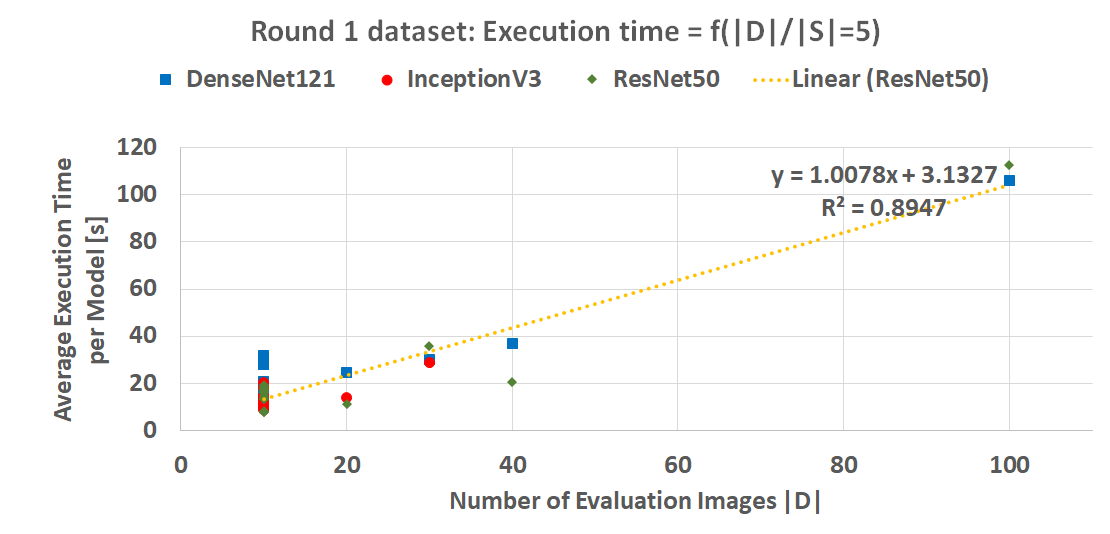}
  \centering
  \caption{Average execution time of model classification for varying numbers of pruned models $|S|$ (top) and evaluated images $|D|$ (bottom). Averages are computed over $nM=1000$ models and for a variety of the six parameters in $\theta_n$. The line is the least squared linear fit to all average execution times for the ResNet50 architecture. }
  \label{fig:09}
\end{figure}

Figure~\ref{fig:09} illustrates the key dependencies of execution time on the number of pruned models $|S|$ and the number of images used for evaluations $|D|$. For fixed $|D|=10$ in Figure~\ref{fig:09} (top), the average of all standard deviations of execution times is 1.45 s for DenseNet121, 1.09 s for InceptionV3, and 1.04 s for ResNet50.
For fixed $|S|=5$ in Figure~\ref{fig:09} (bottom), the average of all standard deviations of execution times is 0.78 s for DenseNet121, 0.81 s for InceptionV3, and 0.61 s for ResNet50. These values indicate that the execution times vary more for the variable $|S|$ than for the variable $|D|$ in our set of explored configurations.

To meet the constraint on $\mathcal{L}_{i}^{exec}$ in Equation \ref{eq:01} for $T_{max}=60$ s, we estimated the values of $|S|\leq15$ and $|D|\leq60$ given our hardware specifications. 
We also observe that the total classification error decreases much faster with increasing $|S|$ ($\approx 0.49\, \:\%$ per $\Delta |S|=1$) than with increasing number of clean evaluation images $|D|$ ($\approx 0.05\,\:\%$ per $\Delta |D|=1$).
The execution times could also be ranked based on NN architectures to ResNet50, InceptionV3, and DenseNet121 from the least to the most time consuming classification.

\underline{Round 2 dataset:}
We evaluated 20 unique pruning configurations applied to 22 model architectures in 165 h of compute time.
The \texttt{Remove}
pruning method could not be applied to three ShuffleNet architectures because of implementation challenges; the ShuffleNet architecture makes it difficult to remove a single filter in grouped convolutions from the dependency graph of pruned modules as input channels and output channels must both be divisible by filter groups \cite{pytorchConv2d2020, groupedConv2d2020}.

The evaluations were applied with the following distribution of parameters:
\begin{equation}
\small
\begin{aligned}
PM &= \{\texttt{Remove} (8x), \texttt{Trim} (7x), \texttt{Reset} (5x)\}\\
SM &= \{\texttt{Targeted} (20x)\}\\
RM &= \{l_1 (19x), \texttt{stdev} (1x)\}\\
p &\in [0.1, 0.4]\\
|S| &= \{5 (12x), 15 (8x)\}\\
|D| &= \{10 (15x), 30 (3x), 40 (1x), All (1x)\}
\end{aligned}
\end{equation}

\begin{figure}
\includegraphics[
  width=8cm,
  keepaspectratio,
]{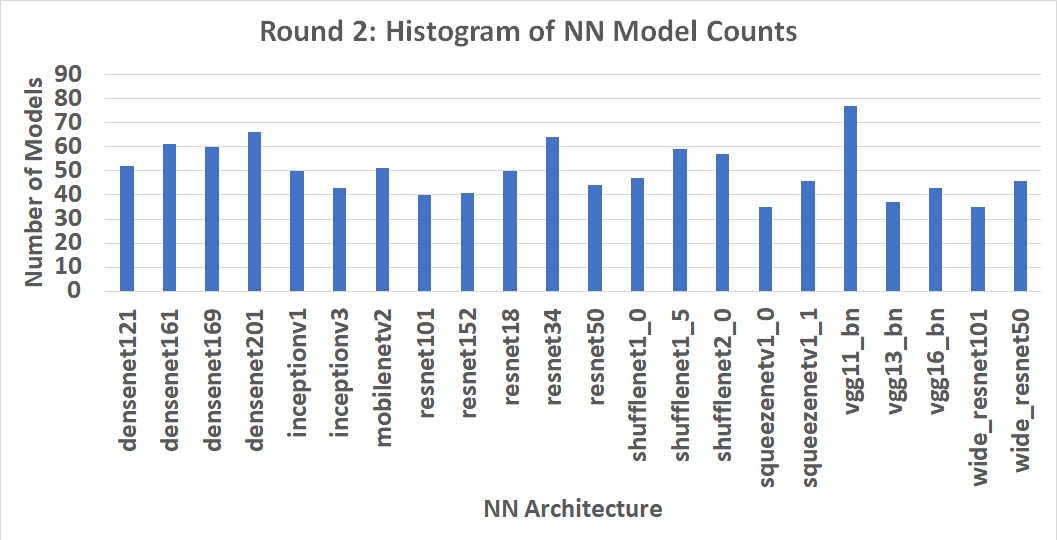}
  \centering
  \caption{Histogram of models architectures in the Round 2 dataset.
  }
  \label{fig:10}
\end{figure}

We mostly evaluated pruning configurations with $SM=\texttt{Targeted}$ and $RM=l_1$ based on the results from the Round 1 dataset analyses. The values of sampling probability $p$ were set according to our assumptions in Section~\ref{sec:methods}. 
Figure \ref{fig:10} shows a histogram of model counts in the Round 2 dataset. Due to the approximately $10\times$ fewer models per architecture than in the Round 1 dataset, the estimate of the mapping  $f(\Vec{A_i}{(M_i, \theta_n)}) \rightarrow L_i$ has likely a larger margin of error.

\begin{figure}
\includegraphics[
  width=8cm,
  keepaspectratio,
]{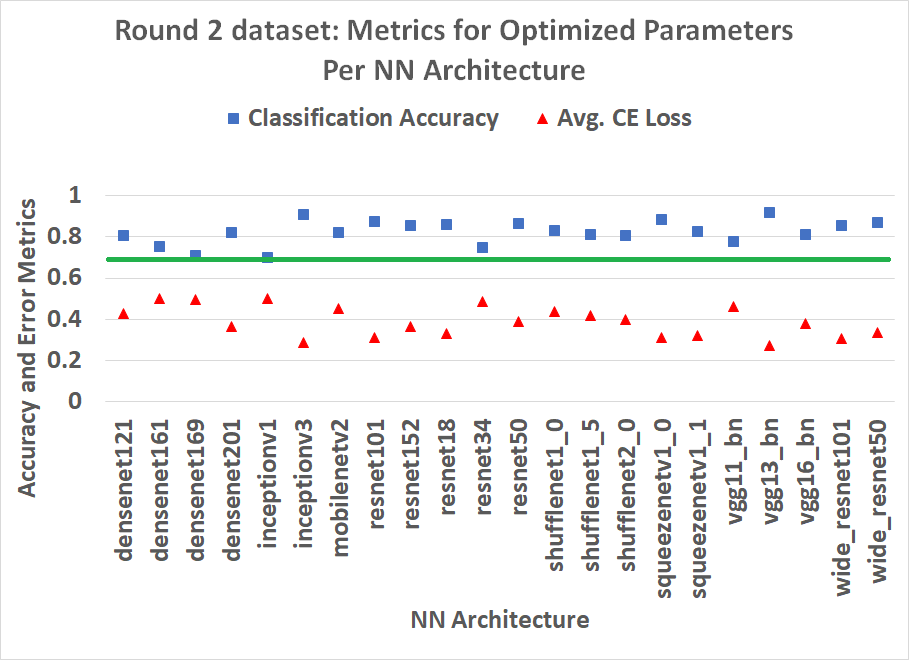}
  \centering
  \caption{Classification accuracy and average cross entropy loss metrics applied to Round 2 dataset for the found optimal parameters $\theta_n$ per architecture. The line at $0.6931$ corresponds to random guessing for CE loss values. }
  \label{fig:12}
\end{figure}

Figure \ref{fig:12} shows the classification accuracy ($\mathcal{L}_{i}^{AC}$) and average CE loss for all found optimal parameters $\theta_n$ per architecture. 
The average classification error over all 22 architectures for the found optimal configurations is $17.59\:\%$ (False Positive = $8.88\:\%$ and False Negative = $8.71\:\%$) and the average CE loss is $0.3888$
(compared to random guessing value $0.6931$). The trojan detection algorithm meets the accuracy requirements \cite{trojai-data2020} set to CE loss = $0.3465$ 
for 8 out of 22 architectures (i.e., InceptionV3, ResNet18, ResNet101, SqueezeNet1.0, SqueezeNet1.1, VGG 13, Wide ResNet50, and Wide ResNet101). The average execution time per model is $41$ s. The execution limit of $60$ s is met by 19 out 22 model architectures (i.e., it is not met by DenseNet169, DenseNet201, and Wide ResNet101).

The parameter values over the optimal settings found confirm that larger $|S|$ values improve detection accuracy. The choice of a pruning method appears to be specific to the architecture, i.e., $|S|=\{5 (1x), 15 (21x)\}$ and $PM=\{\texttt{Remove} (12x), \texttt{Trim} (6x), \texttt{Reset} (4x)\}$ in the 22 optimal configurations $\theta_n$.

\section{Conclusion}

We presented a baseline pruning-based approach to trojan detection that was evaluated on 2104 NN models from TrojAI Challenge (Round 1 and Round 2 datasets). The approach achieved average classification accuracy of $69.73\:\%$  over Round 1 dataset and $82.41\:\%$ over Round 2 dataset with an average processing time of less than $60$ s per model on a CPU hardware. 
The code for such experimentations is available in GitHub~\cite{Bajcsy2021}.


  \section*{Acknowledgments}
    The funding for all authors was provided by IARPA: IARPA-20001-D2020-2007180011


\section*{Disclaimer}
Commercial products are identified in this document in order to specify the experimental procedure adequately.
Such identification is not intended to imply recommendation or endorsement by the National Institute
of Standards and Technology, nor is it intended to imply that the products identified are necessarily the best available for the purpose.


%
%
%

\bibliography{baseline_trojan_detection}

\begin{thebibliography}{10}
\expandafter\ifx\csname url\endcsname\relax
  \def\url#1{\texttt{#1}}\fi
\expandafter\ifx\csname urlprefix\endcsname\relax\def\urlprefix{URL }\fi
\expandafter\ifx\csname href\endcsname\relax
  \def\href#1#2{#2} \def\path#1{#1}\fi

\bibitem{Bajcsy2020}
P.~Bajcsy, N.~Schaub, M.~Majurski, {Scientific Calculator for Designing Trojan
  Detectors in Neural Networks}, Association for the Advancement of Artificial
  Intelligence (AAAI), Fall Symposium Series (FSS), AI in Government and Public
  Sector Applications (2020) 8.

\bibitem{IARPA2020}
IARPA, {Intelligence Advanced Research Projects Agency: Trojans in Artificial
  Intelligence (TrojAI)}, \url{https://pages.nist.gov/trojai/} (1 2020).

\bibitem{relevantPapers2020}
T.~Kulp-McDowall, A.~Dima, M.~Majurski, {TrojAI Literature Review.},
  \url{https://github.com/usnistgov/trojai-literature} (12 2020).

\bibitem{Siegelmann2019}
H.~Siegelmann, {Guaranteeing AI Robustness against Deception (GARD)},
  \url{https://www.darpa.mil/program/guaranteeing-ai-robustness-against-deception}
  (2019).

\bibitem{Xu2019}
X.~Xu, Q.~Wang, H.~Li, N.~Borisov, C.~A. Gunter, B.~Li, {Detecting AI Trojans
  Using Meta Neural Analysis} (2019).

\bibitem{Jha2019}
S.~Jha, S.~Raj, S.~L. Fernandes, S.~K. Jha, S.~Jha, B.~Jalaian, G.~Verma,
  A.~Swami, {Attribution-based confidence metric for deep neural networks},
  Advances in Neural Information Processing Systems 32~(NeurIPS) (2019).

\bibitem{Erichson2020}
N.~B. Erichson, D.~Taylor, Q.~Wu, M.~W. Mahoney, {Noise-response analysis for
  rapid detection of backdoors in deep neural networks}, arXiv (2020).
\newblock \href {http://arxiv.org/abs/2008.00123} {\path{arXiv:2008.00123}}.

\bibitem{baseline2020}
E.~Ameisen, {Always start with a stupid model, no exceptions.},
  \url{https://blog.insightdatascience.com/always-start-with-a-stupid-model-no-exceptions-3a22314b9aaa}
  (3 2018).

\bibitem{Blalock2020}
D.~Blalock, J.~J.~G. Ortiz, J.~Frankle, J.~Guttag, {What is the state of neural
  network pruning?}, arXiv (2020).
\newblock \href {http://arxiv.org/abs/2003.03033} {\path{arXiv:2003.03033}}.

\bibitem{BabakHassibi1992}
B.~Hassibi, D.~G. Stork, {Second Order Derivatives for Network Pruning: Optimal
  Brain Surgeon}, in: Advances in Neural Information Processing Systems 5 (NIPS
  1992), Neural Information Processing Systems Foundation, Inc., 1992, pp.
  164--172.

\bibitem{Han2015}
S.~Han, J.~Pool, J.~Tran, W.~J. Dally, {Learning both Weights and Connections
  for Efficient Neural Networks} (2015).
\newblock \href {http://arxiv.org/abs/1506.02626} {\path{arXiv:1506.02626}}.

\bibitem{Hu2016}
H.~Hu, R.~Peng, Y.-w. Tai, S.~G. Limited, C.-k. Tang, {Network Trimming: A
  Data-Driven Neuron Pruning Approach towards Efficient Deep Architectures}
  (2016).
\newblock \href {http://arxiv.org/abs/1607.03250} {\path{arXiv:1607.03250}}.

\bibitem{Li2017}
H.~Li, A.~Kadav, I.~Durdanovic, H.~Samet, H.~P. Graf, {Pruning Filters for
  Efficient ConvNets}, in: International Conference on Learning
  Representations, Palais des Congr{\`{e}}s Neptune, Toulon, France, 2017, pp.
  1--13.

\bibitem{Liu2018a}
K.~Liu, B.~Dolan-Gavitt, S.~Garg, {Fine-pruning: Defending against backdooring
  attacks on deep neural networks}, Lecture Notes in Computer Science
  (including subseries Lecture Notes in Artificial Intelligence and Lecture
  Notes in Bioinformatics) 11050 LNCS (2018) 273--294.
\newblock \href {https://doi.org/10.1007/978-3-030-00470-5-13}
  {\path{doi:10.1007/978-3-030-00470-5-13}}.

\bibitem{anwar2015structured}
S.~Anwar, K.~Hwang, W.~Sung, Structured pruning of deep convolutional neural
  networks (2015).
\newblock \href {http://arxiv.org/abs/1512.08571} {\path{arXiv:1512.08571}}.

\bibitem{See2016}
A.~See, M.~T. Luong, C.~D. Manning, {Compression of neural machine translation
  models via pruning}, CoNLL 2016 - 20th SIGNLL Conference on Computational
  Natural Language Learning, Proceedings (2016) 291--301\href
  {https://doi.org/10.18653/v1/k16-1029} {\path{doi:10.18653/v1/k16-1029}}.

\bibitem{mariet2017diversity}
Z.~Mariet, S.~Sra, Diversity networks: Neural network compression using
  determinantal point processes (2017).
\newblock \href {http://arxiv.org/abs/1511.05077} {\path{arXiv:1511.05077}}.

\bibitem{Molchanov2017}
P.~Molchanov, S.~Tyree, T.~Karras, T.~Aila, J.~Kautz, Pruning convolutional
  neural networks for resource efficient inference, 5th International
  Conference on Learning Representations, ICLR 2017 - Conference Track
  Proceedings~(2015) (2017) 1--17.
\newblock \href {http://arxiv.org/abs/1611.06440} {\path{arXiv:1611.06440}}.

\bibitem{Ye2018}
J.~Ye, X.~Lu, Z.~Lin, J.~Z. Wang, {Rethinking the smaller-norm-less-informative
  assumption in channel pruning of convolution layers}, arXiv~(2017) (2018)
  1--11.
\newblock \href {http://arxiv.org/abs/1802.00124} {\path{arXiv:1802.00124}}.

\bibitem{Liu2018}
Y.~Liu, S.~Ma, Y.~Aafer, W.-C. Lee, J.~Zhai, W.~Wang, X.~Zhang, {Trojaning
  Attack on Neural Networks}, in: NDSS, Internet Society, Network and
  Distributed Systems Security (NDSS) Symposium 2018, San Diego, CA, 2018, pp.
  1--15.
\newblock \href {https://doi.org/10.14722/ndss.2018.23291}
  {\path{doi:10.14722/ndss.2018.23291}}.

\bibitem{pruninglink2020}
jacobgil, wanglouis49, zepx, eeric, insomnia250, {Model Pruning Implementations
  in GitHub by the listed GitHub users}, \url{https://github.com} (12 2020).

\bibitem{trojai-data2020}
NIST, {Datasets for Trojans in Artificial Intelligence (TrojAI)},
  \url{https://pages.nist.gov/trojai/} (12 2020).

\bibitem{Bajcsy2021}
P.~Bajcsy, {Implementations of Pruning-Based Trojan Detection in GitHub},
  \url{https://github.com/usnistgov/trojai-baseline-pruning} (1 2021).

\bibitem{pickle2020}
Python, Software, Foundation, {Pickle - Python object serialization},
  \url{https://docs.python.org/3/library/index.html} (12 2020).

\bibitem{graphvis2020}
J.~Ellson, E.~Gansner, Y.~Hu, S.~North, {Graphviz - Graph Visualization
  Software}, \url{https://graphviz.org/} (12 2020).

\bibitem{pytorchConv2d2020}
Python, Software, Foundation, {PyTorch Conv2d Class},
  \url{https://pytorch.org/docs/stable/generated/torch.nn.Conv2d.html} (12
  2020).

\bibitem{groupedConv2d2020}
VainF-GitHub, {Grouped Convolution Issue},
  \url{https://github.com/VainF/Torch-Pruning/issues/9} (12 2020).

\end{thebibliography}

%
%
%

\end{document}